\documentclass[showpacs,amsmath,amssymb,prd,nofootinbib,twocolumn]{revtex4}
\pdfoutput=1
\usepackage{graphicx}
\usepackage{dcolumn}
\usepackage{bm}
\usepackage{epsf}
\usepackage{hyperref}
\usepackage{amsfonts}
\usepackage{verbatim}
 
\begin{document}
 
\title{Invariants for Tendex and Vortex Fields}

\author{Kenneth A. Dennison}

\affiliation{Department of Physics and Astronomy, Bowdoin College,
Brunswick, Maine 04011}

\author{Thomas W. Baumgarte}

\affiliation{Department of Physics and Astronomy, Bowdoin College,
Brunswick, Maine 04011}



\begin{abstract}
Tendex and vortex fields, defined by the eigenvectors and eigenvalues of the electric and magnetic parts of the Weyl curvature tensor, form the basis of a recently developed approach to visualizing spacetime curvature.   In analogy to electric and magnetic fields, these fields are coordinate-dependent.  However, in a further analogy, we can form invariants from the tendex and vortex fields that are invariant under coordinate transformations, just as certain combinations of the electric and magnetic fields are invariant under coordinate transformations.  We derive these invariants, and provide a simple, analytical demonstration for non-spherically symmetric slices of a Schwarzschild spacetime.
\end{abstract}


\pacs{04.25.dg, 04.25.Nx, 04.70.Bw, 97.60.Lf}
 
\maketitle

 
Tendex and vortex fields have recently been introduced as tools for the visualization and interpretation of gravitational fields and spacetime dynamics (\cite{OweBCKLMNSZZT11}, see also \cite{Tho12}).  In particular, it has been suggested that they can provide deeper insight into phenomena observed in numerical simulations of binary black hole mergers, including orbital hang-up (see, e.g., \cite{CamLZ06b}), spin-flip (e.g.~\cite{CamLZKM07}) and black hole recoil (e.g.~\cite{HerHHSL07,BakCCKMM06,CamLZM07,GonHSBH07,LouZ11}).  Tendex fields are defined in terms of the eigenvalues and eigenvectors of the electric part of the Weyl curvature tensor and describe tidal stretching or compression, while vortex fields are defined in terms of the eigenvalues and eigenvectors of the magnetic part of the Weyl tensor and describe precession.   Examples of tendex and vortex fields for a number of different spacetimes can be found in references \cite{OweBCKLMNSZZT11,NicOZZBCKLMST11,ZimNZ11,DenB12}. 

In analogy to their electromagnetic counterparts, tendex and vortex fields are not invariant under general coordinate transformations.  However, just as for electromagnetic fields, it is possible to find combinations of the tendex and vortex fields that {\em are} invariant.  In this short paper we derive these invariants and also provide a simple example that illustrates both the coordinate dependence of the tendex and vortex fields themselves, as well as the coordinate-independence of these invariants.

Before considering tendex and vortex fields it is useful to review the properties of electromagnetic fields under coordinate transformations.   Recall that the electric field $E^i$ and the magnetic field $B^i$ are given by components of the four-dimensional, antisymmetric Faraday tensor $F^{ab}$ (here and in the following, indices $i$, $j$, etc.~denote spatial indices, while indices $a$, $b$, etc.~denote spacetime indices).   Under a general coordinate transformation, the Faraday tensor transforms as a rank-2 tensor, and the new electric and magnetic fields can be identified from the Faraday tensor in the new coordinate system.  In particular, a coordinate transformation may therefore mix the electric and magnetic fields.

More specifically, we may write the Faraday tensor as
\begin{equation} \label{faraday}
F^{ab} = n^a E^b - n^b E^a + n_d \epsilon^{dabc} B_c.
\end{equation}
Here we have assumed a foliation of the spacetime $M$ into a family of spatial slices $\Sigma$, each of which corresponds, at least locally, to a level surface of a coordinate time $t$ (see, e.g., \cite{BauS10}).  The normal vector $n^a$ is defined as the future-pointing normal on the spatial slices, and $\epsilon_{abcd}$ is the Levi-Civita tensor.  A normal observer then identifies 
\begin{equation}
\label{EaBadefined}
E^a = F^{ab} n_b,~~~~~~~~~B^a = -\frac{1}{2} \epsilon^{abcd} F_{cd} n_{b}.
\end{equation}
as the electric and magnetic fields.    By construction both vectors are spatial, i.e.~$E^a n_a = B^a n_a = 0$.

Under a general coordinate transformation, level surfaces of the coordinate time $t$ may change, meaning that the new spatial slices may be tilted with respect to the old spatial slices (as is familiar from boosts in special relativity).  This means that the normal vector $n^a$ in the new coordinate system does not represent the same object as that in the old coordinate system -- it points in a different direction.  From the relations (\ref{EaBadefined}) it is therefore clear that $E^a$ and $B^a$ do not transform as four-dimensional rank-1 tensors under general coordinate transformations (even though they do transform as three-dimensional vectors under purely spatial coordinate transformations, which leave $n^a$ unaffected).


However, the electric and magnetic fields inherit invariants from the Faraday tensor.  In particular, from the invariance of the scalar $F_{ab} F^{ab}$ it follows that
\begin{equation} \label{I_EM_1}
{\mathcal I}_1^{\rm EM} \equiv E^2 - B^2
\end{equation}
must be invariant.  Here $E$ and $B$ denote the magnitudes of the electric and magnetic fields.  A second invariant 
\begin{equation} \label{I_EM_2}
{\mathcal I}_2^{\rm EM} \equiv E_a B^a = E_i B^i = E B \cos \theta_{EB}
\end{equation}
can be derived from the invariance of the determinant of the Faraday tensor (under transformations between orthonormal frames).  Here $\theta_{EB}$ denotes the angle between $E^i$ and $B^i$.  It can also be shown that any other invariants for the electromagnetic fields are combinations of these two (see, e.g., Problem 4.2 in \cite{LigPPT75}). 

In general relativity, we may similarly write the Weyl curvature tensor $C_{abcd}$ in terms of its electric part ${\mathcal E}_{ab}$ and its magnetic part ${\mathcal B}_{ab}$,
 \begin{eqnarray}
C_{abcd} &=& 4\left(n_{[a}n_{[c}+\gamma_{[a[c}\right){\mathcal E}_{b]d]} + 2\epsilon_{abe}n_{[c}{\mathcal B}_{d]}{}^{e} \nonumber\\&&+\:2n_{[a}{\mathcal B}_{b]e}\epsilon_{cd}{}^{e},
\end{eqnarray}
where $\epsilon_{abc}=n^{d}\epsilon_{dabc}$ is the spatial Levi-Civita tensor, and where $\gamma_{ab}$ is the spatial metric induced on the spatial slice (see \cite{MaaB98}).  A normal observer then identifies the electric and magnetic parts as
\begin{equation}
{\mathcal E}_{ab} = C_{a c b d}n^{c}n^{d},~~~~
{\mathcal B}_{ab} = - {}^* C_{acbd} n^{c}n^{d}
\end{equation}
where ${}^* C_{abcd} \equiv \frac{1}{2} \epsilon_{ab} {}^{ef} C_{efcd}$ denotes the dual of the Weyl tensor 
(compare \cite{ZhaBSL12}).  By construction, both ${\mathcal E}_{ab}$ and ${\mathcal B}_{ab}$ are spatial, symmetric and traceless.  By the same arguments as above we also note that ${\mathcal E}_{ab}$ and ${\mathcal B}_{ab}$ do not transform as four-dimensional rank-2 tensors.  Their behavior under general coordinate transformations instead follows from that of the rank-4 Weyl tensor $C_{abcd}$. 

In analogy to their electromagnetic counterparts, ${\mathcal E}_{ab}$ and ${\mathcal B}_{ab}$ inherit invariants from the Weyl tensor.  In vacuum, the Weyl tensor has four independent algebraic real invariants, which may be written as the real and imaginary parts of the scalars  
\begin{equation}
{\mathcal I} = \frac{1}{16}\left(C_{abcd}C^{cdab}-{\rm i}C_{abcd}{}^{*}C^{cdab}\right)
\end{equation}
and 
\begin{equation}
{\mathcal J} =\frac{1}{96}\left(C_{abcd}C^{cdef}C_{ef}{}^{ab} - {\rm i}C_{abcd}C^{cdef}{}^{*}C_{ef}{}^{ab}\right),
\end{equation}
(\cite{SteKMHH03}, see also \cite{BeeB02,BeeBBN05}).  These well-known scalars play a central role in the Petrov classification of spacetimes (see, e.g., \cite{SteKMHH03}).  They also form the basis of some spacetime diagnostics, including the specialty index suggested by \cite{BakC00}, and a radiation scalar suggested by \cite{BeeB02}.  In \cite{ZhaBSL12}, they are used to identify a geometrically motivated coordinate system for numerical-relativity simulations.  In terms of  ${\mathcal E}_{ij}$ and ${\mathcal B}_{ij}$, the invariants ${\mathcal I}$ and ${\mathcal J}$ can be written as 
\begin{equation} \label{I}
{\mathcal I} = \frac{1}{2}\left({\mathcal E}_{ij}{\mathcal E}^{ij} - {\mathcal B}_{ij}{\mathcal B}^{ij}\right) + {\rm i}{\mathcal E}_{ij}{\mathcal B}^{ij}
\end{equation}
and
\begin{eqnarray} \label{J}
{\mathcal J} &=& \left(-\frac{1}{6}{\mathcal E}^{i}{}_{j}{\mathcal E}^{j}{}_{k}{\mathcal E}^{k}{}_{i} + \frac{1}{2}{\mathcal E}^{i}{}_{j}{\mathcal B}^{j}{}_{k}{\mathcal B}^{k}{}_{i}\right) \nonumber\\&& +\:{\rm i}\left(\frac{1}{6}{\mathcal B}^{i}{}_{j}{\mathcal B}^{j}{}_{k}{\mathcal B}^{k}{}_{i} - \frac{1}{2}{\mathcal B}^{i}{}_{j}{\mathcal E}^{j}{}_{k}{\mathcal E}^{k}{}_{i}\right).  
\end{eqnarray}
(see \cite{BurBB06}, \cite{ZhaBSL12}).

We now express these invariants in terms of tendex and vortex fields.  Tendex fields are defined by 
the eigenvalues and eigenvectors of the electric part of the Weyl tensor, ${\mathcal E}_{ij}$, while vortex fields are defined by those of the magnetic part ${\mathcal B}_{ij}$.  The eigenvalues $\lambda$ and eigenvectors $v^i$ are most easily found in an orthonormal basis (denoted with hats)
\begin{equation}
{\mathcal E}^{\hat{\imath}}{}_{\hat{\jmath}}v^{\hat{\jmath}} = \lambda v^{\hat{\imath}},
\end{equation}
and similar for ${\mathcal B}_{ij}$. A negative (positive) tendex eigenvalue indicates that observers will tend to be tidally stretched (compressed) in a direction aligned with the corresponding tendex eigenvector, while a positive (negative) vortex eigenvalue indicates that a gyroscope will exhibit (counter)clockwise differential precession relative to a nearby gyroscope in the direction of the corresponding vortex eigenvector \cite{OweBCKLMNSZZT11,NicOZZBCKLMST11}.

Since both ${\mathcal E}_{ab}$ and ${\mathcal B}_{ab}$ are spatial and symmetric, they each have a set of three orthonormal eigenvectors.  We may therefore write the two tensors in terms of their eigenvalues $\lambda$ and orthonormal eigenvectors $v^i$ as 
\begin{eqnarray}
{\mathcal E}^{ij} & = & \lambda_{E1} (v_{E1})^i (v_{E1})^j + \lambda_{E2} (v_{E2})^i (v_{E2})^j 
	\nonumber\\&&+\: \lambda_{E3} (v_{E3})^i (v_{E3})^j \nonumber \\
	& = & \sum_{k = 1}^{3} \lambda_{Ek} (v_{Ek})^i (v_{Ek})^j 
 \end{eqnarray}
 and similarly
\begin{equation}
{\mathcal B}^{ij} = \sum_{k = 1}^{3} \lambda_{Bk} (v_{Bk})^i (v_{Bk})^j.
\end{equation}
We use explicit summations for sums over eigenvalues and eigenvectors, but will continue to use the implied Einstein summation rule for sums over vector indices.

We can now derive invariants for the tendex and vortex fields by inserting these expressions into the invariants ${\mathcal I}$ (eq.~(\ref{I})) and ${\mathcal J}$ (eq.~(\ref{J})).  From the real part of ${\mathcal I}$ we find the first invariant
\begin{equation}
\label{I_GR_1}
{\mathcal I}_{1} = \frac{1}{2}\sum_{i = 1}^{3}\left( \lambda_{Ei}^{2} - \lambda_{Bi}^{2}\right),
\end{equation}
in close analogy to the electromagnetic invariant (\ref{I_EM_1}).  Likewise, the second invariant follows from the imaginary part of ${\mathcal I}$,
\begin{eqnarray}
\label{I_GR_2}
{\mathcal I}_2 &=& \sum_{i = 1}^{3} \sum_{j = 1}^{3} \lambda_{Ei}\lambda_{Bj}\left(\left(v_{Ei}\right)_{k}\left(v_{Bj}\right)^{k}\right)^{2} \nonumber\\&=& \sum_{i = 1}^{3} \sum_{j = 1}^{3} \lambda_{Ei}\lambda_{Bj}\cos^{2}\theta_{EiBj},
\end{eqnarray}  
and closely resembles the electromagnetic invariant (\ref{I_EM_2}).  The last two invariants,
\begin{equation}
\label{J_GR_1}
{\mathcal J}_{1} = -\frac{1}{6}\sum_{i = 1}^{3}\lambda_{Ei}^{3} + \frac{1}{2}\sum_{i = 1}^{3}\sum_{j = 1}^{3}\lambda_{Ei}\lambda_{Bj}^{2}\cos^{2}\theta_{EiBj}
\end{equation}
and
\begin{equation}
\label{J_GR_2}
{\mathcal J}_{2} = \frac{1}{6}\sum_{i = 1}^{3}\lambda_{Bi}^{3} - \frac{1}{2}\sum_{i = 1}^{3}\sum_{j = 1}^{3}\lambda_{Bi}\lambda_{Ej}^{2}\cos^{2}\theta_{EiBj},
\end{equation}
can be derived from the real and imaginary parts of ${\mathcal J}$.  

We note in passing that only three of the nine angles $\theta_{EiBj}$ between $(v_{Ei})^k$ and $(v_{Bj})^k$ are independent.  It is possible to write these nine angles in terms of three Euler angles, but that does not appear particularly useful in this context.


As a simple analytical demonstration we now consider a Schwarzschild spacetime.  On a slice of constant Schwarzschild time $t$, the spatial metric, expressed in Schwarzschild coordinates, is
\begin{equation}
\label{tconstspatialmetric}
\gamma_{ij} = \mbox{diag}\left(\left(1- 2M/R \right)^{-1},R^2,R^2\sin^2\theta\right),
\end{equation}
while the extrinsic curvature $K_{ij}$ vanishes.  In a spherical polar orthonormal basis the electric part of the Weyl tensor is
\begin{equation}
\label{tconstEijSchw}
{\mathcal E}_{\hat{\imath}\hat{\jmath}} = \mbox{diag}\left(-2M/R^{3}, M/R^{3},M/R^{3}\right),
\end{equation}
(see, e.g., \cite{MisTW73,NicOZZBCKLMST11}), so that we can immediately identify the eigenvalues
\begin{equation}
\label{tconstEijorthoevals}
\lambda_{E1}  = -\frac{2M}{R^{3}},~~~~~\lambda_{E2} = \lambda_{E3} = \frac{M}{R^{3}}
\end{equation}
and can choose the corresponding eigenvectors to be
\begin{equation}
\label{tconstEijevecs}
(v_{E1})^{\hat{\imath}} = (e_{\hat R})^{\hat \imath},~~~~~(v_{E2})^{\hat{\imath}} = (e_{\hat \theta})^{\hat \imath},~~~~~(v_{E3})^{\hat{\imath}} = (e_{\hat \phi})^{\hat \imath}.
\end{equation}
The magnetic part of the Weyl tensor ${\mathcal B}_{\hat{\imath}\hat{\jmath}}$ and its associated vortex fields, meanwhile, vanish identically on slices of constant Schwarzschild time.  


Now consider a different slicing of the same spacetime.  The new slices are level surfaces of a new time coordinate $\bar t$, which we constructed from the old time coordinate $t$ with the help of a ``height function" $h$,
\begin{equation}
\bar{t} = t + h.
\end{equation}
We will introduce a coordinate system that breaks spherical symmetry by choosing $h$ to be a function of $\theta$,  $h = h(\theta)$. Transforming to the new coordinate system, we can now 
identify the spatial metric and extrinsic curvature on a slice of constant time $\bar{t}$ (see \cite{BauS10} for a pedagogical example).   The spatial metric is given by equation (\ref{tconstspatialmetric}), but with $\gamma_{\theta\theta}=R^{2}$ replaced with
\begin{equation}
\gamma_{\theta\theta} = R^{2} \eta,
\end{equation}
where we have defined
\begin{equation}
\eta = 1 - \frac{(\partial_\theta h)^2}{R^2}\left(1 - \frac{2M}{R} \right).
\end{equation}
When $h$ is constant, we have $\eta = 1$ and recover results for slices of constant time $t$.  The extrinsic curvature no longer vanishes; its nonzero components are now
\begin{eqnarray}
K_{R\theta} &=& K_{\theta R} = \frac{1}{\eta^{1/2}} \frac{\left(\partial_{\theta}h\right)}{R^2}
\frac{3M-R}{\sqrt{1-2M/R}},\\
K_{\theta\theta} & = & \frac{1}{\eta^{1/2}} \left(\partial^{2}_{\theta}h\right)\sqrt{1-2M/R},
\end{eqnarray}
and
\begin{equation}
K_{\phi\phi} = \frac{1}{\eta^{1/2}} \left(\partial_{\theta}h\right)\sqrt{1-2M/R}\sin\theta\cos\theta.
\end{equation}
In a spherical polar orthonormal basis the electric part of the Weyl tensor is given by equation (\ref{tconstEijSchw}), but with ${\mathcal E}_{\hat{R}\hat{R}}=-2M/R^{3}$ replaced with
\begin{equation}
\label{tbarconstErr}
{\mathcal E}_{\hat{R}\hat{R}} = \frac{M}{R^{3}}\left(1 - \frac{3}{\eta} \right),
\end{equation}
and with ${\mathcal E}_{\hat{\phi}\hat{\phi}} = M/R^{3}$ replaced with
\begin{equation}
\label{tbarconstEphiphi}
{\mathcal E}_{\hat{\phi}\hat{\phi}} = \frac{M}{R^{3}}\left(-2+\frac{3}{\eta} \right).
\end{equation}
Since ${\mathcal E}_{\hat{\imath}\hat{\jmath}}$ is still diagonal, the eigenvectors are still given by equations (\ref{tconstEijevecs}), and the corresponding eigenvalues are
\begin{equation}
\lambda_{E1} = {\mathcal E}_{\hat{R}\hat{R}},~~~~~\lambda_{E2} = {\mathcal E}_{\hat{\theta}\hat{\theta}} = \frac{M}{R^{3}},~~~~~\lambda_{E3} = {\mathcal E}_{\hat{\phi}\hat{\phi}},   
\end{equation}
with ${\mathcal E}_{\hat{R}\hat{R}}$ and ${\mathcal E}_{\hat{\phi}\hat{\phi}}$ given by equations (\ref{tbarconstErr}) and (\ref{tbarconstEphiphi}), respectively.

For a slice of constant time $\bar{t}$, the only non-vanishing components of the magnetic part of the Weyl tensor are 
\begin{equation}
{\mathcal B}_{\hat{R}\hat{\phi}} = {\mathcal B}_{\hat{\phi}\hat{R}} = 
\frac{3M}{R^{3}}\frac{\sqrt{1-\eta}}{\eta}.
\end{equation}
The eigenvectors of ${\mathcal B}_{\hat{\imath}\hat{\jmath}}$ are
\begin{equation}
(v_{B1})^{\hat{\imath}} = \frac{1}{\sqrt{2}}\left((e_{\hat{R}})^{\hat{\imath}}+(e_{\hat{\phi}})^{\hat{\imath}}\right),
\end{equation}
\begin{equation}
(v_{B2})^{\hat{\imath}} = (e_{\hat{\theta}})^{\hat{\imath}},
\end{equation}
and
\begin{equation}
(v_{B3})^{\hat{\imath}} =\frac{1}{\sqrt{2}}\left((e_{\hat{R}})^{\hat{\imath}}-(e_{\hat{\phi}})^{\hat{\imath}}\right),
\end{equation}
with eigenvalues
\begin{equation}
\label{Bijtbareval1}
\lambda_{B1} =  - \lambda_{B3} = {\mathcal B}_{\hat{R}\hat{\phi}} = 
\frac{3M}{R^{3}}\frac{\sqrt{1-\eta}}{\eta}
\end{equation}
and
\begin{equation}
\label{Bijtbareval2}
\lambda_{B2} = 0.
\end{equation}
This example highlights the fact that tendex and vortex fields are slicing-dependent concepts.

We are now ready to calculate the invariants (\ref{I_GR_1})-(\ref{J_GR_2}) in both coordinate systems.  It is easiest to start with a slice of constant Schwarzschild time $t$.  In that case, ${\mathcal B}_{\hat{\imath}\hat{\jmath}}$ vanishes, so its eigenvalues and eigenvectors do as well.  Substituting equations (\ref{tconstEijorthoevals}) into equations (\ref{I_GR_1})-(\ref{J_GR_2}), we readily find
\begin{equation}
{\mathcal I}_{1} = \frac{3M^{2}}{R^{6}},~~~~~{\mathcal J}_{1} = \frac{M^{3}}{R^{9}},~~~~~{\mathcal I}_{2} ={\mathcal J}_{2} = 0. 
\end{equation}

On the slice of constant coordinate time $\bar t$ we have $\lambda_{B2}=0$, so that the invariant
${\mathcal I}_{2}$ reduces to
\begin{equation}
{\mathcal I}_{2} = \sum^{3}_{i=1}\lambda_{Ei}\left(\lambda_{B1}\cos^{2}\theta_{EiB1} + \lambda_{B3}\cos^{2}\theta_{EiB3}\right).
\end{equation}
But we have $\cos^{2}\theta_{EiB3}=\cos^{2}\theta_{EiB1}$ for each eigenvector $(v_{Ei})^k$, as well as $\lambda_{B3}=-\lambda_{B1}$, so that ${\mathcal I}_2$ again vanishes.  The argument for ${\mathcal J}_{2}$ is very similar.

To evaluate the invariant ${\mathcal I}_{1}$ on a $\bar t = const$ slice we first observe that
\begin{equation}
\sum^{3}_{i=1}\lambda_{Ei}^{2} = \frac{6M^{2}}{R^{6}}+\frac{18M^{2}}{R^{6}}\frac{1-\eta}{\eta^{2}}.
\end{equation}  
Adding the contributions of the eigenvalues $\lambda_{Bi}^2$ from equations (\ref{Bijtbareval1}) and (\ref{Bijtbareval2}) we see that the second term in the above expression cancels exactly, leaving
${\mathcal I}_{1} = 3M^{2}/R^{6}$ as before.  

To calculate ${\mathcal J}_{1}$, we note that the first term of (\ref{J_GR_1}) evaluates to
\begin{equation}
\label{J_GR_1_part}
-\frac{1}{6}\sum_{i=1}^{3}\lambda_{Ei}^{3} = \frac{M^3}{R^9}\left(1+\frac{9\left(1-\eta\right)}{2\eta^{2}}\right).
\end{equation} 
Carrying out the sum over $j$ in the second term of (\ref{J_GR_1}), and using $\lambda_{B2} = 0$, we find
\begin{eqnarray}
\label{J_GR_1_part2}
\frac{1}{2}\sum_{i = 1}^{3}\sum_{j = 1}^{3}\lambda_{Ei}\lambda_{Bj}^{2}\cos^{2}\theta_{EiBj}=\frac{9}{2}\frac{M^{2}}{R^6}\frac{1-\eta}{\eta^{2}}\nonumber\\\times\:\sum_{i=1}^{3}\lambda_{Ei}\left(\cos^{2}\theta_{EiB1}+\cos^{2}\theta_{EiB3}\right).
\end{eqnarray}
After considering the angles between the eigenvectors and using the fact that the electric part of the Weyl tensor is traceless, we see that (\ref{J_GR_1_part2}) can be written as
\begin{equation}
\frac{1}{2}\sum_{i = 1}^{3}\sum_{j = 1}^{3}\lambda_{Ei}\lambda_{Bj}^{2}\cos^{2}\theta_{EiBj}=-\frac{M^3}{R^9}\frac{9\left(1-\eta\right)}{2\eta^{2}}.
\end{equation}
Adding this to equation (\ref{J_GR_1_part}) shows that the invariant ${\mathcal J}_1$ remains ${\mathcal J}_{1} = M^{3}/R^{9}$, which is not surprising but reassuring.

In conclusion, we have derived slicing-invariant quantities for tendex and vortex fields from well-known scalar curvature invariants.  These invariants play the same role as the invariants for electric and magnetic fields in electromagnetism.  We also demonstrate both the coordinate-dependent nature of the tendex and vortex fields, as well as the coordinate-independence of our new invariants, for non-spherical slices of a Schwarzschild spacetime.

 

\acknowledgments

This work was supported in part by NSF Grant PHY-1063240 to Bowdoin College.


\end{document}